\documentclass[useAMS,usenatbib]{mn2e}

\usepackage{amssymb}
\usepackage{amsfonts}
\usepackage{mncite}
\usepackage{epsfig}
\usepackage{psfig}

\newcommand{\de}{\mbox{d}}

\newcommand{\msunyr}{M_{\sun}/\mbox{yr}}

\defcitealias{LR04}{Paper I}

\begin{document}

\title[Locality of transport in self-gravitating discs  - II]{Testing the
  locality of transport in self-gravitating accretion discs - II. The
  massive disc case}

\author[G. Lodato \& W.K.M. Rice]{G. Lodato$^1$ and W.K.M. Rice$^2$ \\
$^1$ Institute of Astronomy, Madingley Road, Cambridge, CB3 0HA\\
$^2$ Institute of Geophysics and Planetary Physics and Department of 
  Earth Sciences, University of California, Riverside, CA 92521} 

\maketitle

\begin{abstract}

In this paper, we extend our previous analysis \citep{LR04} of the
transport properties induced by gravitational instabilities in
cooling, gaseous accretion discs to the case where the disc mass is
comparable to the central object. In order to do so, we have performed
global, three-dimensional smoothed particle hydrodynamics simulations
of massive discs. These new simulations show a much more complex
temporal evolution with respect to the less massive case. Whereas in
the low disc mass case a self-regulated, marginally stable state
(characterized by an approximately constant radial profile of the
stability parameter $Q$) is easily established, in the high disc mass
case we observe the development of an initial transient and subsequent
settling down in a self-regulated state in some simulations, or a
series or recurrent spiral episodes, with low azimuthal wave number
$m$, in others. Accretion in this last case can therefore be a highly
variable process. On the other hand, we find that the secular
evolution of the disc is relatively slow. In fact, the time-average of
the stress induced by self-gravity results in accretion time-scales
much longer than the dynamical timescale, in contrast with previous
isothermal simulations of massive accretion discs. We have also
compared the resulting stress tensor with the expectations based on a
local theory of transport, finding no significant evidence for global
wave energy transport.

\end{abstract}


\section{Introduction}

It is now becoming progressively more clear (both from the theoretical
and from the observational point of view) that gravitational
instabilities might play an important role in the dynamics of
accretion discs, both on the large scales of Active Galactic Nuclei
(AGN) and on the small scales of protostellar and protoplanetary
discs.

At the AGN scale, a powerful observational tool is provided by water
maser observations of the gas motions at distances of $\approx 1$ pc
from the central black hole \citep{greenhill97,kondratko04}, which
often show that the rotation curve can depart significantly from a
Keplerian profile, suggesting that the disc itself might give a
significant contribution to the gravitational field. Indeed, in the
case of the Seyfert galaxy NGC 1068, simple self-gravitating disc
models are able to reproduce the observed non-Keplerian rotation curve
\citep{LB03a}.

The case for self-gravitating accretion discs is maybe even stronger in
the context of star and planet formation. It has long been argued
\citep{linpringle87} that in the early stages of star formation the
disc mass can be sufficiently high for the effect of its self-gravity
to become important. Simple estimates by \citet{larson84} lead to the
conclusion that a gravitationally unstable disc is a likely outcome of
the isothermal collapse of protostellar clouds.  The recent discovery
of a 100 $M_{\odot}$ disc surrounding a massive (20 $M_{\odot}$)
protostar in the M17 nebula \citep{chini04} would clearly indicate, if
the mass estimates are confirmed, that the formation of a massive star
undergoes a phase of self-gravitating disc accretion. In the case of
low mass (proto)stars, disc mass estimates are uncertain, but discs in
the upper end of the mass distribution for T Tauri stars
\citep{launhardt2001,natta04} can have a mass $\approx 30$\% of the
central star. In addition, there have also been theoretical suggestions
that the disc self-gravity might be important in the outer disc of FU
Orionis objects \citep{bellin94,armitage2001,LB03b}.

Recently, the disc self-gravity has been considered especially in view
of the possibility of forming smaller mass companions by the direct
fragmentation of a gravitationally unstable disc, with obvious
implications for the formation of gas giant planets in a protostellar
disc (see, for example, \citealt{boss02,rice03b} or, for similar
arguments on the galactic scale, \citealt{goodman04}). However, it has
now become clear that the outcome of gravitational instabilities
strongly depends on the thermodynamics of the disc. In particular, the
fragmentation of a gravitationally unstable disc requires that the disc
be able to cool very efficiently \citep{gammie01,rice03a}, with a
cooling rate $t_{\mathrm{cool}} \lesssim 3\Omega^{-1}$ (where $\Omega$
is the angular velocity of the disc). This is a rather severe
requirement for a protostellar disc, so that the formation of planets
from the direct fragmentation of a massive disc seems rather unlikely
(however, the inclusion of the effects of convection might
substantially reduce the cooling timescale, see \citealt{boss04a}). In
any case, even the simple development of a gravitational instability in
the form of a spiral structure can significantly accelerate the process
of planet formation by favouring the agglomeration of planetesimals
\citep{haghighipour03a,rice04}. Moreover, gravitational stresses might
play another important role in the process of planet formation, since
the mixing induced by these stresses can influence the early evolution
of chemical elements in protoplanetary discs \citep{boss04b}.

Another important issue related to the disc self-gravity is its ability
to transport angular momentum and thus to promote the process of
accretion. This has been recognized in the analysis of isothermal
simulations of massive discs \citep{laughlin94,laughlin96} and
subsequently analysed in a number of papers
\citep{pickett98,pickett2000,nelson2000}. However, only recently have
simulations been performed which include a relatively more complex
treatment of the thermal status of the disc
\citep{rice03a,pickett03,mejia04}. In a previous paper (\citealt{LR04},
hereafter Paper I), we have explored in detail the transport properties
of self-gravitating, cooling discs, by performing some high-resolution
smoothed particle hydrodynamics (SPH) simulations. These simulations
were used to explore the extent to which angular momentum and energy
transport via gravitational instability can be regarded as a local
phenomenon, or whether the intrinsic global nature of the instabilities
would preclude such an approach. The simulations described in
\citetalias{LR04} included disc masses up to $0.25M_{\star}$ (where
$M_{\star}$ is the mass of the central object) and showed how, when the
disc is allowed to cool down due to external cooling and to heat up as
a consequence of the development of gravitational spiral instabilities,
it eventually settles down in a marginally stable state, characterized
by an approximately flat profile of the parameter $Q$:
\begin{equation}
\label{eq:q}
Q=\frac{c_{\mathrm{s}}\kappa}{\pi G\Sigma}\approx 1,
\end{equation}
where $c_{\mathrm{s}}$ is the thermal speed, $\kappa$ is the epicyclic
frequency and $\Sigma$ is the surface density of the disc. The
resulting spiral structure turned out to be a quasi-stationary
structure, evolving only on the long ``viscous'' time-scale, and able
to transport angular momentum and energy at a rate which is in
reasonable agreement with the expectations based on a local (viscous)
theory of transport.

It can be shown that, for self-regulated discs, a simple relationship
holds between the disc aspect ratio $H/R$ and the mass ratio
$M_{\mathrm{disc}}/M_*$. The disc is in fact marginally stable when
its temperature is small enough that $H/R\lesssim
M_{\mathrm{disc}}/M_*$. Of course, this relationship can only hold in
the limit where $M_{\mathrm{disc}}/M_*\ll 1$. As
$M_{\mathrm{disc}}/M_*$ grows, the disc becomes thicker, and global
effects are going to be more important. This was already shown by us
in Paper I. The limit $M_{\mathrm{disc}}/M_*\sim 1$ is therefore
particularly interesting, because we expect a different physics to
come into play.

In this paper we extend our previous analysis, considering the case
where the disc mass is particularly high. We have in fact considered
the two cases where $M_{\mathrm{disc}} = 0.5 M_{\star}$ and
$M_{\mathrm{disc}} = 1M_{\star}$. The results of these new simulations
present significant differences with respect to our previous
analysis. In the high disc mass case, we now find the development of
transient, global spiral structures, which last for roughly one
dynamical timescale and develop recurrently during the course of the
simulation. While in the $M_{\mathrm{disc}} = 0.5 M_{\star}$ case,
after a first transient, the disc is eventually able to settle down in
a quasi-stationary state, in the equal mass case the disc in not able
to ``self-regulate'' the strength of the spiral disturbance in order
to counterbalance exactly the externally imposed cooling and
short-lived global disturbances recurrently develop during the whole
simulation. Accretion in this last case will therefore appear to be a
highly variable process.

The paper is organized as follows. In Section \ref{sec:basic} we
briefly review the relevant physical concepts of the problem and the
details of the numerical setup that we adopt. In Section
\ref{sec:results} we describe the results of our simulations. In
Section \ref{sec:conclusions} we discuss our results and draw our
conclusions.

\section{Basic concepts and numerical setup}
\label{sec:basic}

\subsection{Basic physical quantities}

In this section we briefly review the basic concepts about the
evolution of viscous discs and the development of gravitational
instabilities and their associated transport properties. More details
can be found in section 2 of \citetalias{LR04} and references therein.

The equations of motion for an axisymmetric disc, in
cylindrical coordinates, are the equation of continuity:
\begin{equation}
\label{continuity}
\frac{\partial\Sigma}{\partial t}+\frac{1}{R}\frac{\partial}{\partial
R} (R\Sigma u) =0,
\end{equation}
and the azimuthal component of Euler's equation (expressed in terms of
angular momentum conservation):
\begin{equation}
\label{angmomcons}
\frac{\partial}{\partial t}(\Sigma
R^2\Omega)+\frac{1}{R}\frac{\partial} {\partial R} (\Sigma
R^3\Omega u)=-\frac{1}{R}\frac{\partial}{\partial R}(R^2 T_{R\phi}),
\end{equation}
where $u$ is the radial velocity and $T_{R\phi}$ is the relevant
component of the viscous stress tensor, integrated in the vertical
direction. This last term is the basic ingredient in the theory of
accretion discs. Standard hydrodynamical viscosity is not sufficient
to provide accretion at the required rates, and therefore $T_{R\phi}$
is generally described by means of {\it ad hoc} prescriptions.  The
$\alpha$-prescription \citep{shakura73}, based on simple arguments on
the relevant physical scales of turbulent cells in the discs, assumes
that $T_{R\phi}$ is proportional to the disc pressure:
\begin{equation}
T_{R\phi}=\left|\frac{\mbox{d}\ln\Omega}{\mbox{d}\ln
  R}\right|\alpha\Sigma c_s^2,
\label{alpha}
\end{equation}
where the proportionality constant $\alpha$ is an unknown parameter,
usually considered to be smaller than unity. The $\alpha$-prescription
can be also put in the following equivalent form, which involves the
kinematical viscosity coefficient $\nu$:
\begin{equation}
\label{eq:alphashakura}
\nu=\alpha c_{\mathrm s} H,
\end{equation}
where $H$ is the thickness of the disc.

The effect of viscosity on the energy balance is twofold: viscous
torques convect energy between neighbouring annuli of the disc and they
dissipate energy. The energy which is convected across a ring at
radius $R$ per unit time is given by:
\begin{equation}
\label{eq:entransport}
\frac{\de E}{\de t}=2\pi R^2 T_{R\phi}\Omega,
\end{equation}
while the dissipation rate per unit surface is given by:
\begin{equation}
\label{eq:dissipation}
D(R)=T_{R\phi} |R\Omega'|.
\end{equation}
If the disc is in thermal equilibrium, we can derive a useful relation
between the viscosity coefficient $\alpha$ and the cooling
timescale. If we assume that cooling can be simply parameterised in
the following way:
\begin{equation}
\label{cooling}
Q^-=\frac{U}{t_{\mathrm{cool}}}=\frac{\Sigma c_s^2}{\gamma(\gamma-1)
t_{\mathrm{cool}}},
\end{equation}
where $U$ is the internal energy per unit surface and $\gamma$ is the
ratio of the specific heats, then, equating the viscous dissipation
term, expressed in Eq.  (\ref{eq:dissipation}) to the cooling term,
expressed in Eq. (\ref{cooling}), leads to:
\begin{equation}
\label{eq:gammie}
\alpha=\left|\frac{\de\ln\Omega}{\de\ln R}\right|^{-2}\frac{1} 
{\gamma(\gamma-1)t_{\mathrm{cool}}\Omega},
\end{equation}
where we have used also Eq. (\ref{alpha}).

\subsection{Diagnostics for gravitationally induced transport}

In order to describe the development and the properties of a
self-gravitating structure, we will use the same diagnostics adopted in
\citetalias{LR04}, and we summarize them here.

The basic parameter that describes the local, axisymmetric stability of
a gaseous disc with respect to gravitational instabilities is the $Q$
parameter defined in Eq. (\ref{eq:q}). It actually turns out that the
stability criterion based on this parameter, namely that $Q\gtrsim1$,
is quite robust, and generally applicable also in the case of global,
non-axisymmetric disturbances (see also Paper I). As mentioned in the
Introduction, and as discussed in more details elsewhere (Paper I,
\citealt{bertin97,pacinski78}) the competitive effects of external
cooling and effective heating due to gravitational instabilities tend
to result in a self-regulated state, where the value of $Q$ is constant
throughout the disc and very close to its marginal stability value. The
simulations presented in \citetalias{LR04} confirmed the effectiveness
of the self-regulation mechanism. We will therefore track the
development of a self-regulated state by looking at the profile of $Q$.

The relevant component of the stress tensor associated with
gravitational instabilities is defined as \citep{lyndenbell72}:
\begin{equation}
T_{\mathrm{R\phi}}^{\mathrm{grav}}=\int\de
z\frac{g_{\mathrm{R}}g_{\phi}}{4\pi G},
\end{equation}
where $g_{\mathrm{R}}$ and $g_{\phi}$ are the radial and azimuthal
component of the gravitational field, respectively. The full stress
tensor also includes the 'Reynolds' stress, defined as:
\begin{equation}
T_{\mathrm{R\phi}}^{\mathrm{Reyn}}=\Sigma\delta v_{\mathrm{R}}
\delta v_{\mathrm{\phi}},
\end{equation}
where $\delta{\mathbf v}=\mathbf{v-u}$, with $\mathbf{v}$ the fluid
velocity and $\mathbf{u}$ the mean fluid velocity. It is always
possible to define an effective $\alpha$ parameter associated with
gravitational instabilities, by setting:
\begin{equation}
T_{\mathrm{R\phi}}^{\mathrm{grav}}+T_{\mathrm{R\phi}}^{\mathrm{Reyn}} =
\left|\frac{\de\ln\Omega}{\de\ln R}\right|\alpha\Sigma
c_{\mathrm{s}}^2.
\label{eq:torque_alpha}
\end{equation}
A local description of angular momentum transport then requires that
the stress tensor, and therefore $\alpha$, at a given location is only
determined by local conditions in the disc. In Paper I we have already
shown that in self-gravitating discs a region of size $\approx 10H$
contributes to the stress at a given location, therefore implying that
for discs as thick as $H/R\approx 0.1$ the stress tensor is not
determined locally. 

In addition, there is also another issue concerning the use of a
viscous formalism in self-gravitating accretion discs. In fact,
\citet{balbus99} have shown that in general the energy transport
provided by gravitational instabilities contains global terms,
associated with wave energy transport, which cannot be directly
associated with an effective viscosity. It remains to be discussed how
large these terms are, and under which conditions they play a
significant role. \citet{balbus99} argue that global energy transport
should not play an important role in a marginally stable disc. In this
case, a viscous description of transport should be appropriate and
therefore energy should be transported according to
Eq. (\ref{eq:entransport}) and it should be dissipated according to
Eq. (\ref{eq:dissipation}). If the disc is in thermal equilibrium, the
latter statement is equivalent to the prescription that $\alpha$ and
$t_{\mathrm{cool}}$ are related through Eq. (\ref{eq:gammie}). This
offers us a simple test to check whether energy dissipation is
actually viscous or not in massive discs. In particular, since in all
our simulations (as discussed below), we adopt $\gamma=5/3$ and
$t_{\mathrm{cool}}=7.5\Omega^{-1}$, the expected value of $\alpha$ for
viscous dissipation to balance cooling is $\alpha\approx 0.053$.

\subsection{Numerical issues and setup}

The numerical setup is largely identical to that used in Paper I. We
use Smoothed Particle Hydrodynamics (SPH) (see \citealt{benz90,
monaghan92}) to perform three-dimensional simulations of gaseous
accretion discs. We consider a system comprising a central star, with
mass $M_*$, surrounded by a disc, with mass $M_{\rm disc}$, and we
consider mass ratios, $q = M_{\rm disc}/M_*$, of $0.5$ and $1$. The
disc is modelled using 250000 SPH particles while the central object is
modelled as a point mass onto which gas particles may accrete if they
move to within the accretion radius. Both the gas particles and the
point mass use a tree to determine neighbours and to calculate
gravitational forces \citep{benz90}, and the central object is free to
move under the gravitational influence of the disc gas.

The disc particles are distributed such as to give an initial surface
density profile of $\Sigma \propto R^{-1}$ with an initial temperature
profile chosen to be $T \propto R^{-1/2}$. The initial velocity profile
is calculated by including the enclosed mass and the pressure gradient
when determining the angular frequency $\Omega$. With these initial
conditions, the minimum value of $Q$ is attained at the outer edge of
the disc. For each simulation, the temperature normalisation is chosen
such that the minimum value of $Q$ is $Q_{\rm min} = 2$, giving a disc
that is initially gravitationally stable. The disc is also assumed to
be in vertical hydrostatic equilibrium (see, for example,
\citealt{pringle81}).  The particles are therefore distributed such
that the vertical density profile is a Gaussian with a scaleheight $H$
given by $H=c_{\mathrm{s}}/\Omega$. Actually, in a self-gravitating
disc, the vertical density profile is not rigorously Gaussian
\citep{BL99,hure2000}, so our initial setup, strictly speaking, is not
in dynamical equilibrium. However, dynamical equilibrium is achieved
rapidly (i.e., in a dynamical timescale) during the simulation.

Our calculations are essentially scale free. In dimensionless units,
the disc extends from $R_{\rm in} = 0.25$ to $R_{\rm out} = 25$ and the
star has a mass of $M_* = 1$. In these units, the orbital period at $R
= 1$ is $2 \pi$ code units and one orbital period at the outer edge of
the disc is approximately 800 time units.

Since the main aim of this work, as in Paper I, is to investigate
transport properties of self-gravitating accretion discs, we explicitly
solve the energy balance for the gas. We allow the disc to heat up due
to both $P\de V$ work and viscous dissipation and assume that the ratio of
the specific heats is $\gamma = 5/3$. We then impose a cooling of the
form \citep{gammie01}
\begin{equation}
\label{cooling2}
\frac{\de u_i}{\de t}=-\frac{u_i}{t_{\mathrm{cool}}},
\end{equation}
where $u_i$ is the internal energy per unit mass and
$t_{\mathrm{cool}}$ is given by $t_{\mathrm{cool}} = \beta
\Omega^{-1}$. Previous simulations \citep{gammie01, rice03c} have shown
that discs fragment for $\beta \le 3$ with this fragmentation boundary
possibly increasing as the disc mass increases \citep{rice03c}. To
ensure that our discs do not fragment, we use $\beta = 7.5$. The
imposed cooling should lead to the growth of the gravitational
instability which will heat the disc, returning it to a state of
marginal stability. There will also be additional heating through the
artificial SPH viscosity. In these simulation we use the standard SPH
viscosity (e.g. \citealt{monaghan92}) but also include the Balsara
switch \citep{balsara95} to reduce the shear component of the
artificial viscosity. As discussed in detail in Paper I, the angular
momentum transport induced by the artificial viscosity is at least a
factor of 10 smaller than that due to the gravitational
instabilities. The artificial viscosity should therefore play only a
minor role in the disc dynamics.

\begin{figure*}
\centerline{ \epsfig{figure=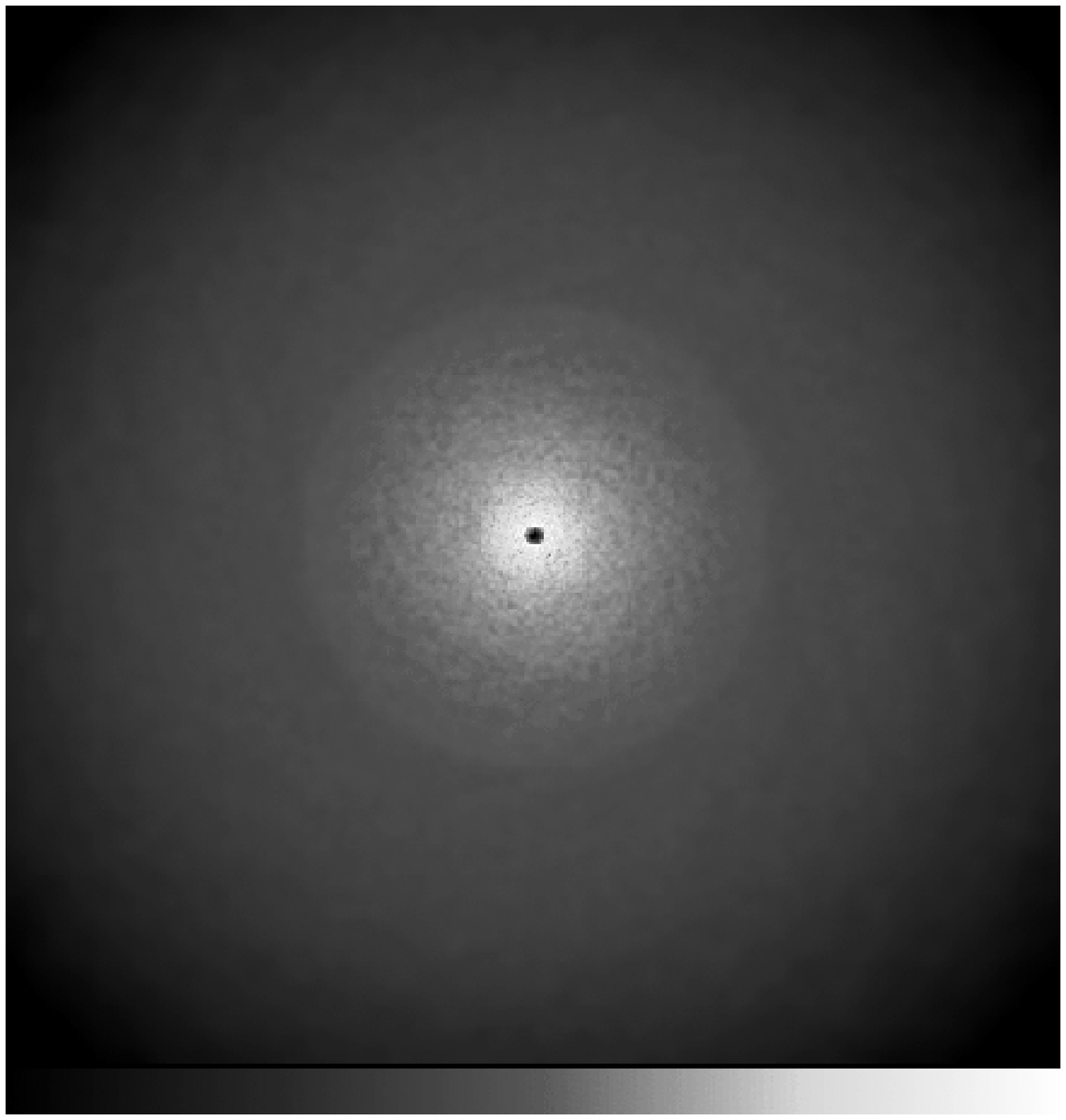,width=.45\textwidth}
             \epsfig{figure=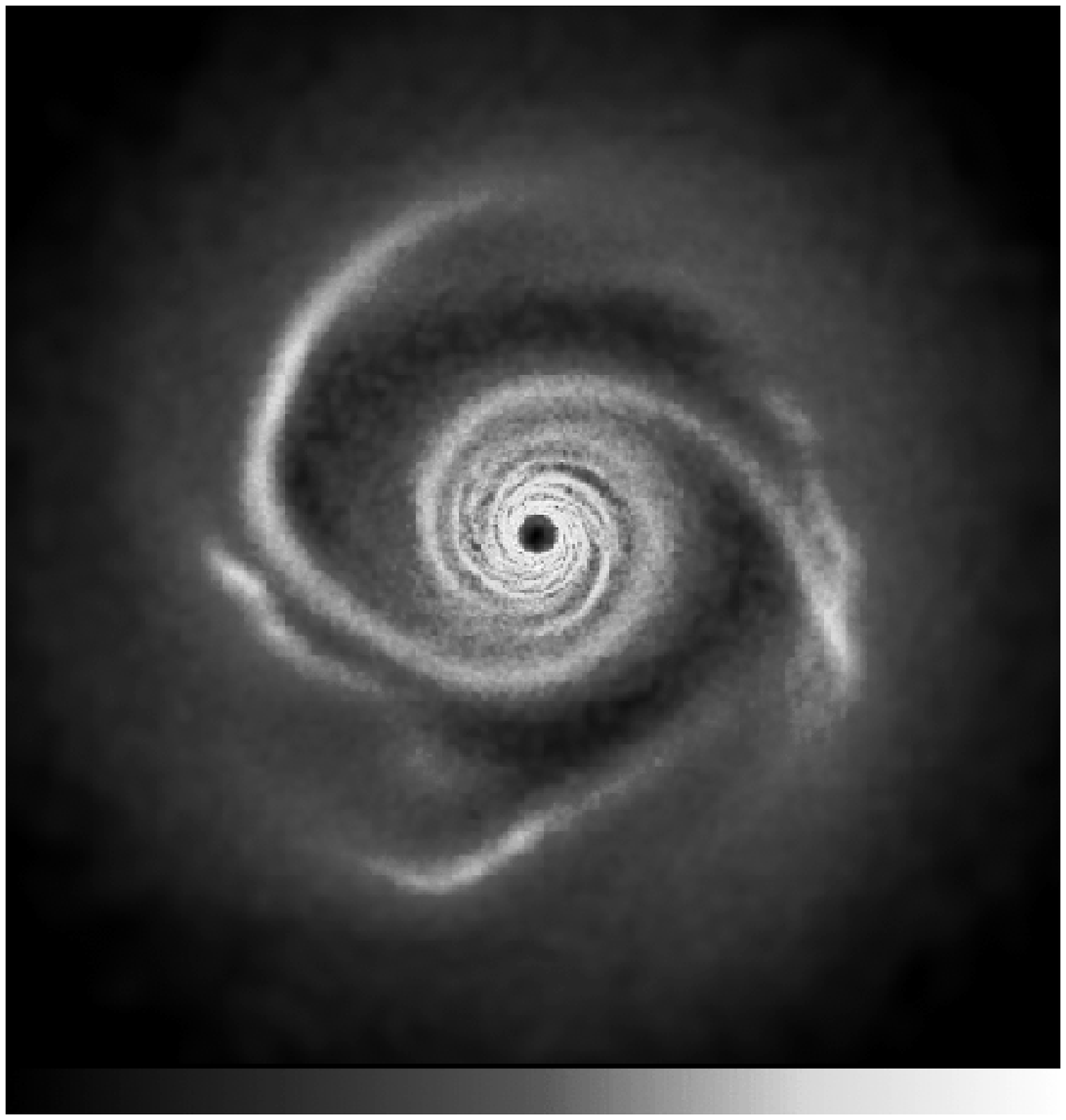,width=.45\textwidth}}
\centerline{ \epsfig{figure=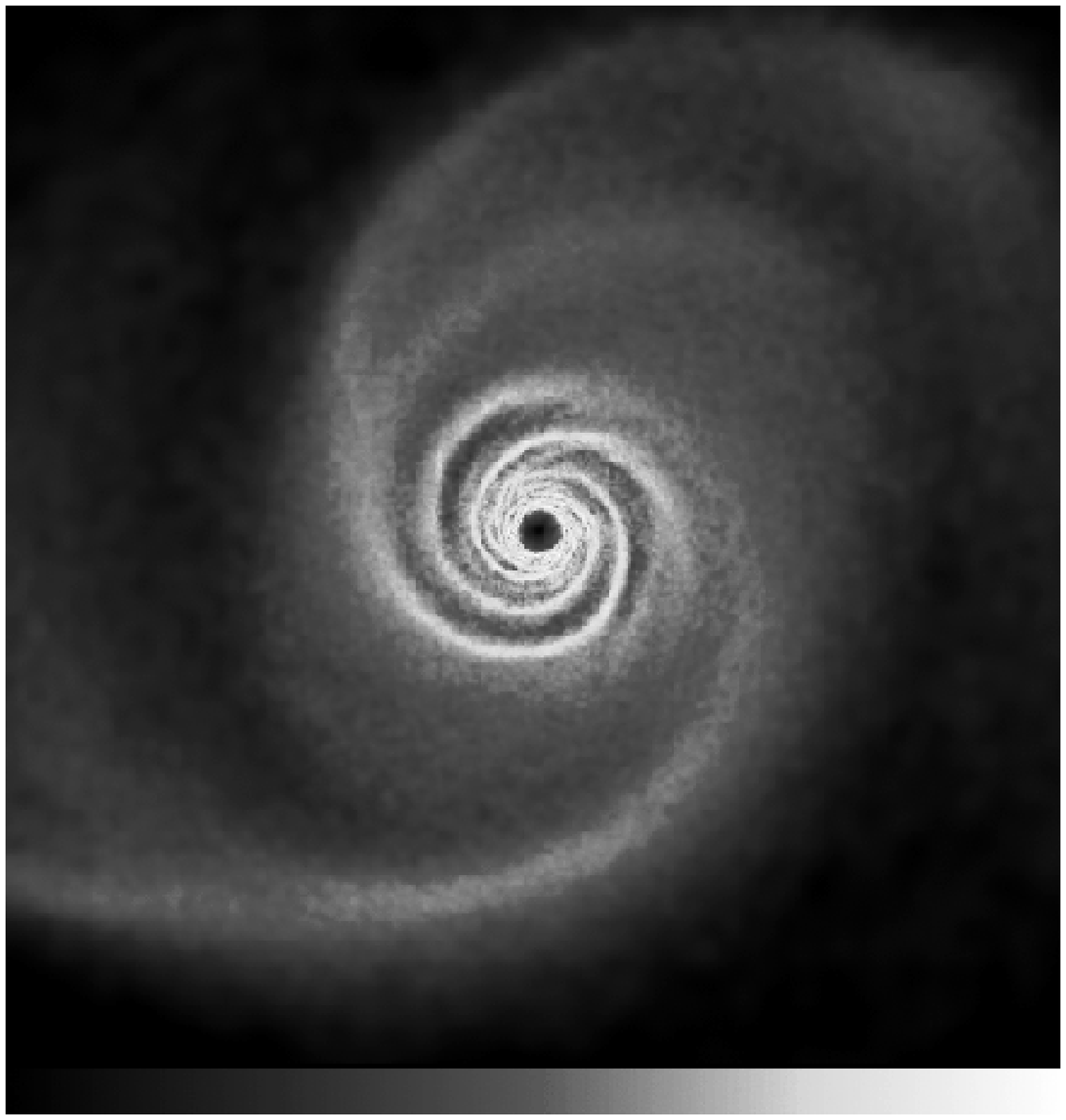,width=.45\textwidth}
             \epsfig{figure=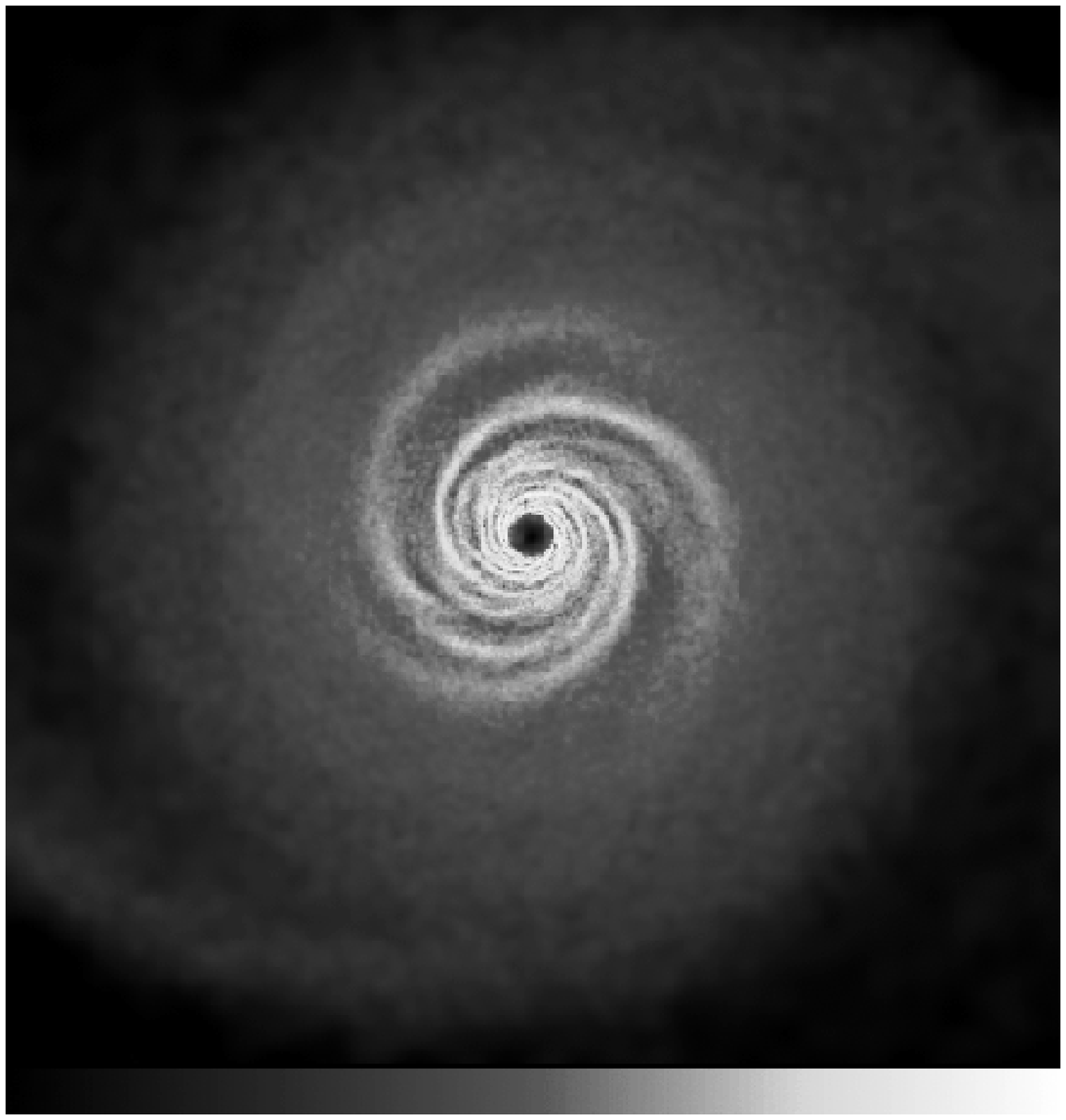,width=.45\textwidth}}
\caption{Time evolution of the logarithm of the surface density
$\Sigma$ (with a colour scale ranging between $10^4-10^7$ g cm$^-2$)
during the development of the transient spiral structure at the
beginning of the $M_{\mathrm{disc}}=0.5M_{\star}$ simulation. They
refer to $t=0$ (upper left), $t\approx 1.9 t_0$ (upper right),
$t\approx 2.5 t_0$ (lower left) and $t\approx 3t_0$ (lower right). The
linear scale of the images ranges from -25 to 25 (in code units) for
both axes.}
\label{fig:transient}
\end{figure*}

\section{Results}
\label{sec:results}

In this section we describe the main results of our simulations. As
mentioned above, all the simulations are scale-free and can therefore
describe the basic dynamics of different astrophysical systems (from
protostellar discs, to AGN discs or even to galactic discs), even if we
do not intend to model the detailed physical appearance of a specific
one. However, when numbers are needed, we will consider as a reference
case, for illustrative purposes, a ``low mass proto-star circumstellar
disc'', where the central object mass is $1M_{\odot}$ and the unit
length of the simulations is $R_0=1$ au. Thus, in this reference case,
the disc extends from $R_{\mathrm{in}}=0.25$ au to
$R_{\mathrm{out}}=25$ au. We will express all times in units of the
{\it Keplerian} rotation period of the outer disc
$t_0=2\pi/\Omega_{\mathrm{out}}$, where
$\Omega_{\mathrm{out}}^2=GM_{\star}/R_{\mathrm{out}}^3$. For most
cases, we have followed the simulations up to $\approx 7t_0$, i.e. up
to a few thermal timescales at the outer edge of the disc.

\subsection{The $M_{\mathrm{disc}}=0.5M_{\star}$ case}

The evolution of the disc in this case is characterized by two phases:
({\it i}) soon after the beginning of the simulation the disc develops
a large-scale, transient spiral structure that subsequently vanishes,
({\it ii}) at later times the disc settles down in a quasi-stationary
state, characterized by a much slower evolution. Here we describe the
two stages separately.

\subsubsection{The initial transient}

In the earliest stages of the simulation the disc simply cools down,
until (at roughly $t=2t_0$) the minimum value of $Q$ approaches unity
and the first gravitational disturbances start to develop. These are
dominated by a grand-design, $m=2$ spiral mode. The pattern frequency
of this mode (i.e. the angular velocity of a reference frame in which
the spiral is stationary) is $\Omega_{\mathrm{P}}\approx 1.5
\Omega_{\mathrm{out}}$. In this case the corotation radius (i.e. the
radius at which the spiral mode and the gas rotate with the same
angular velocity) is $R\approx 19$, very close to the outer edge of the
disc. This large-scale structure is however not long-lived, and has
essentially disappeared at $t\approx 3t_0$ (i.e. after one dynamical
timescale), leaving behind a smaller-scale disturbance characterized by
modes with larger $m$ and confined within the inner disc. Fig.
\ref{fig:transient} shows the logarithm of the disc surface density
(with a colour scale ranging between $10^4-10^7$ g cm$^{-2}$) during the
development of this transient disturbance. The four images refer to
$t=0, 1.9, 2.5$ and $3t_0$, respectively. The linear scale of the
images ranges between -25 and 25 for both axes.

\begin{figure}
\centerline{\epsfig{figure=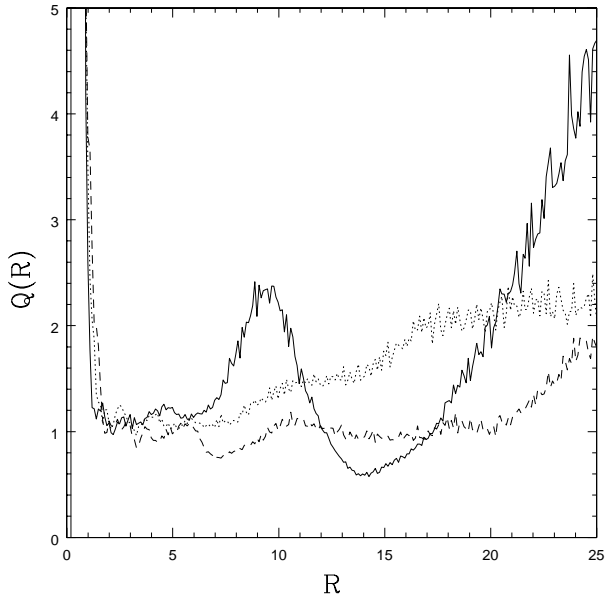,width=84mm}}
\caption{Evolution of the profile of $Q$ during the transient in the
  $M_{\mathrm{disc}}=0.5M_{\star}$ simulation. The three curves refer
  to $t=1.9t_0$ (solid line), $t=2.7t_0$ (dotted line) and $t=7t_0$
  (dashed line).}
\label{fig:Qtrans}
\end{figure}

The development of this spiral is accompanied by a significant
modification of the profile of $Q$. Fig. \ref{fig:Qtrans} shows the
$Q$-profile at $t=1.9t_0$ (solid line), $t=2.7t_0$ (dotted line) and,
for comparison, at $t=7t_0$ (i.e. at the end of the simulation, when a
self-regulated state is eventually achieved; dashed line). It can be
seen that this first spiral transient leads to a strong ``heating'' of
the outer disc, with $Q$ rising well above unity. In this way the outer
disc is stabilized and the large-scale spiral structure rapidly
vanishes.

Another important feature associated with this transient spiral
instability is the mass transport induced by it. We have evaluated the
strength of the torques excited by the gravitational instability as
discussed in section \ref{sec:basic}. The left panel of
Fig. \ref{fig:alphatrans} shows the evolution of the parameter $\alpha$
at four different times during the transient: $t=2.28t_0$ (solid line),
$t=2.52t_0$ (dashed line), $t=2.69t_0$ (long-dashed line) and
$t=2.93t_0$ (dot-dashed line). The two dotted lines show the expected
``equilibrium'' value $\alpha=0.053$, obtained from balancing viscous
heating and cooling (as discussed above in section \ref{sec:basic}),
and the maximum torque strength ($\alpha=0.005$) due to artificial SPH
viscosity (see Paper I, Appendix). It is apparent that the transient
induces a strong angular momentum transport in the outer disc, with
maximum values of $\alpha$ raising above 0.2. At later times, when the
transient vanishes, the value of $\alpha$ decreases accordingly,
becoming much smaller than 0.053 in the outer disc, and close to 0.053
within $R=5$. The right panel of Fig. \ref{fig:alphatrans} shows the
corresponding mass accretion rates (if we use the scale units of the
reference case). The $\dot{M}$ induced by the transient is significant
in the outer disc, becoming as large as $\approx 10^{-4} \msunyr$, and
then decreasing significantly once the transient has disappeared.

\begin{figure}
\centerline{\epsfig{figure=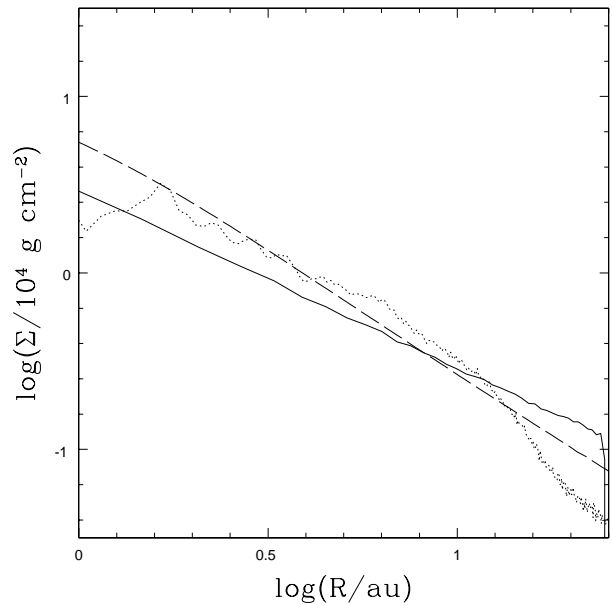,width=84mm}}
\caption{Azimuthally averages surface density profiles at $t=0$ (solid
line) and at $t=2.93t_0$ (dotted line). The dashed line shows the
surface density profile predicted by the self-regulated disc models of
\citet{BL99}.}
\label{fig:sigmatrans}
\end{figure}

The large mass accretion rate of course induces a significant
redistribution of matter in the disc. This is shown in
Fig. \ref{fig:sigmatrans}, where we plot the azimuthally averaged
surface density of the disc at $t=0$ (solid line) and at $t=2.93t_0$
(dotted line). The spiral structure has produced a steepening of the
surface density profile. 

\citet{BL99} have studied a class of steady-state, self-regulated
massive accretion disc models. The surface density profile of these
models asymptotically approaches $\Sigma \sim r^{-1}$ at large radii,
but it shows significant departures from a simple power-law at small
radii, becoming progressively steeper. These self-regulated models
depend on only one dimensional parameter: a scale-length
$R_\mathrm{s}$, defined as follows:
\begin{equation}
\label{eq:rs}
R_{\mathrm{s}}=2GM_{\star}\left(\frac{\bar{Q}}{4}\right)^2
\left(\frac{G\dot{M}}{3\alpha}\right)^{-2/3},
\end{equation}
where $\bar{Q}\approx 1$ is the marginally stable value of $Q$. The
dashed line in Fig. \ref{fig:sigmatrans} shows the surface density
profile of this analytical model, with $R_\mathrm{s}=200$ au. With this
choice, the mass enclosed within 25 au is $0.5M_{\star}$, i.e. the same
as in our simulation. This density profile matches well the surface
density profile obtained after the occurrence of the transient
spiral. This suggest that this episode of gravitational instability is
the response of the disc to the initial conditions assumed here. The
disc's ``preferred'' state is a self-regulated state, in which the
surface density profile is slightly steeper than the initial
$\Sigma\propto R^{-1}$. The large-scale spiral transient that we
observe has the effect of moving just enough matter so as to steepen
the surface density profile to match the one implied by the
self-regulation condition.

\subsubsection{Long-term evolution}

\begin{figure}
\centerline{\epsfig{figure=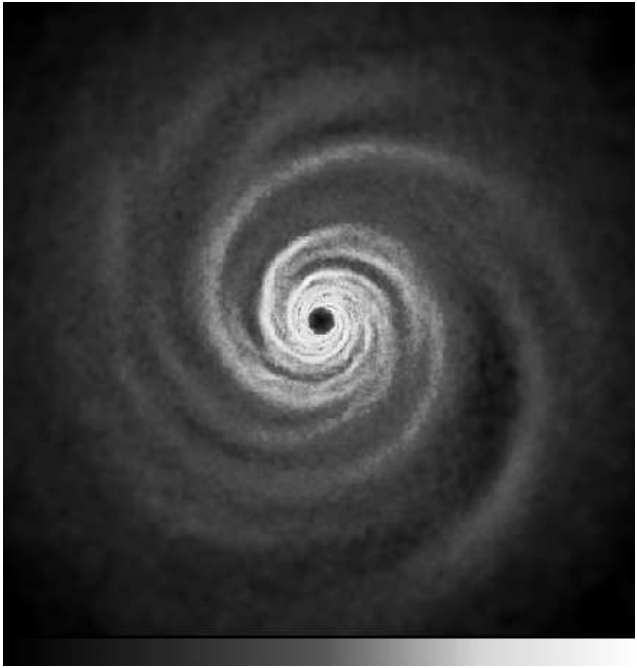,width=84mm}}
\caption{Surface density in the $M_{\mathrm{disc}} =0.5M_{\star}$ case,
at the end of the simulation. The scales are the same as in
Fig. \ref{fig:transient}.}
\label{fig:md05image}
\end{figure}

After the first transient, the value of $Q$ in the outer disc is
relatively high (see Fig. \ref{fig:Qtrans}) and the spiral structure is
confined to the inner disc. In this way, the main source of heating in
the outer disc (i.e., the effective heating due to gravitational
instabilities) is turned off and the disc cools down until a new
generation of gravitational disturbances eventually
develops. Fig. \ref{fig:md05image} shows the surface density of the
disc at the end of the simulation. The scales are exactly the same as
in Fig. \ref{fig:transient}.

\begin{figure*}
\centerline{\epsfig{figure=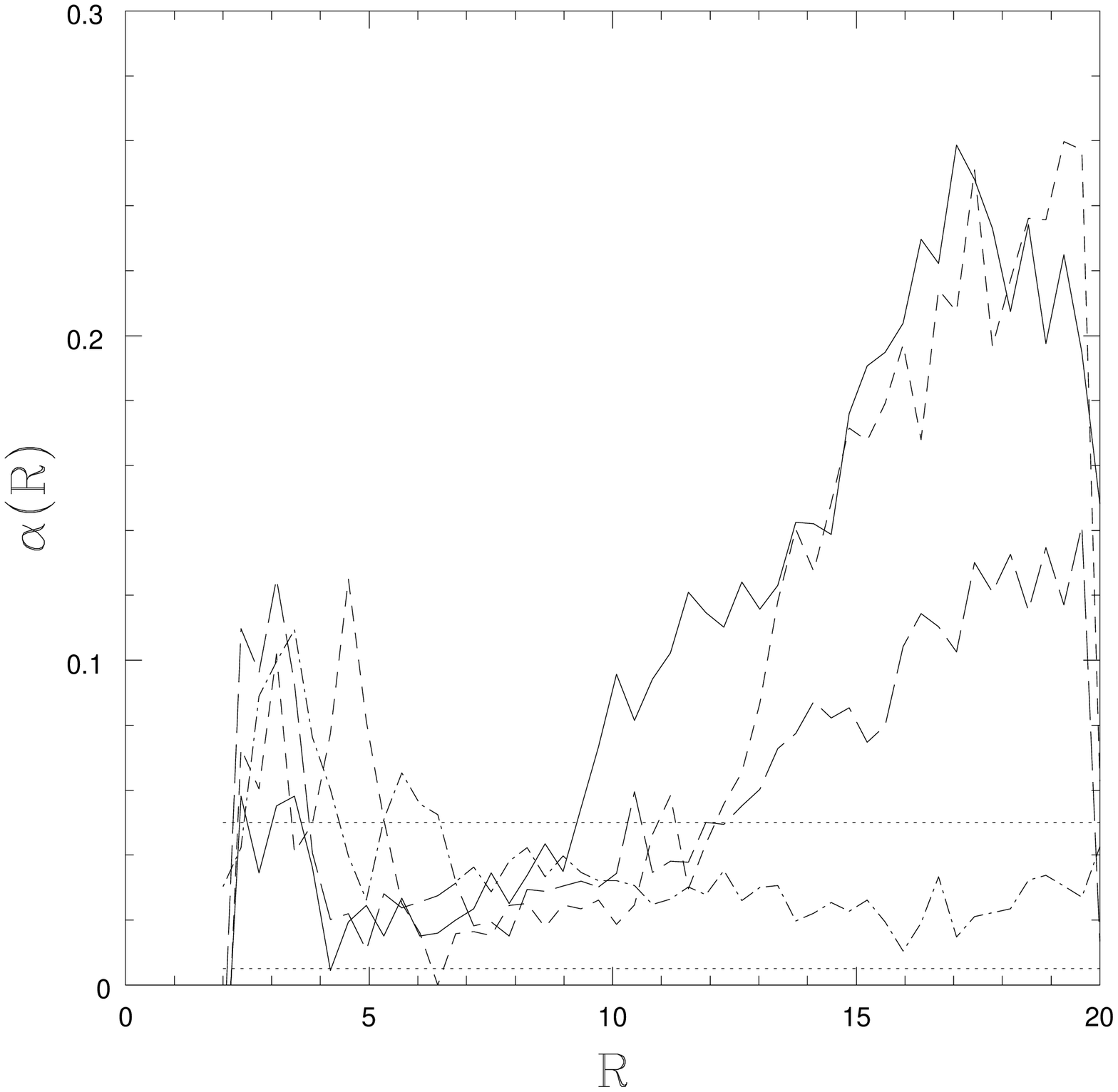,width=.45\textwidth}
            \epsfig{figure=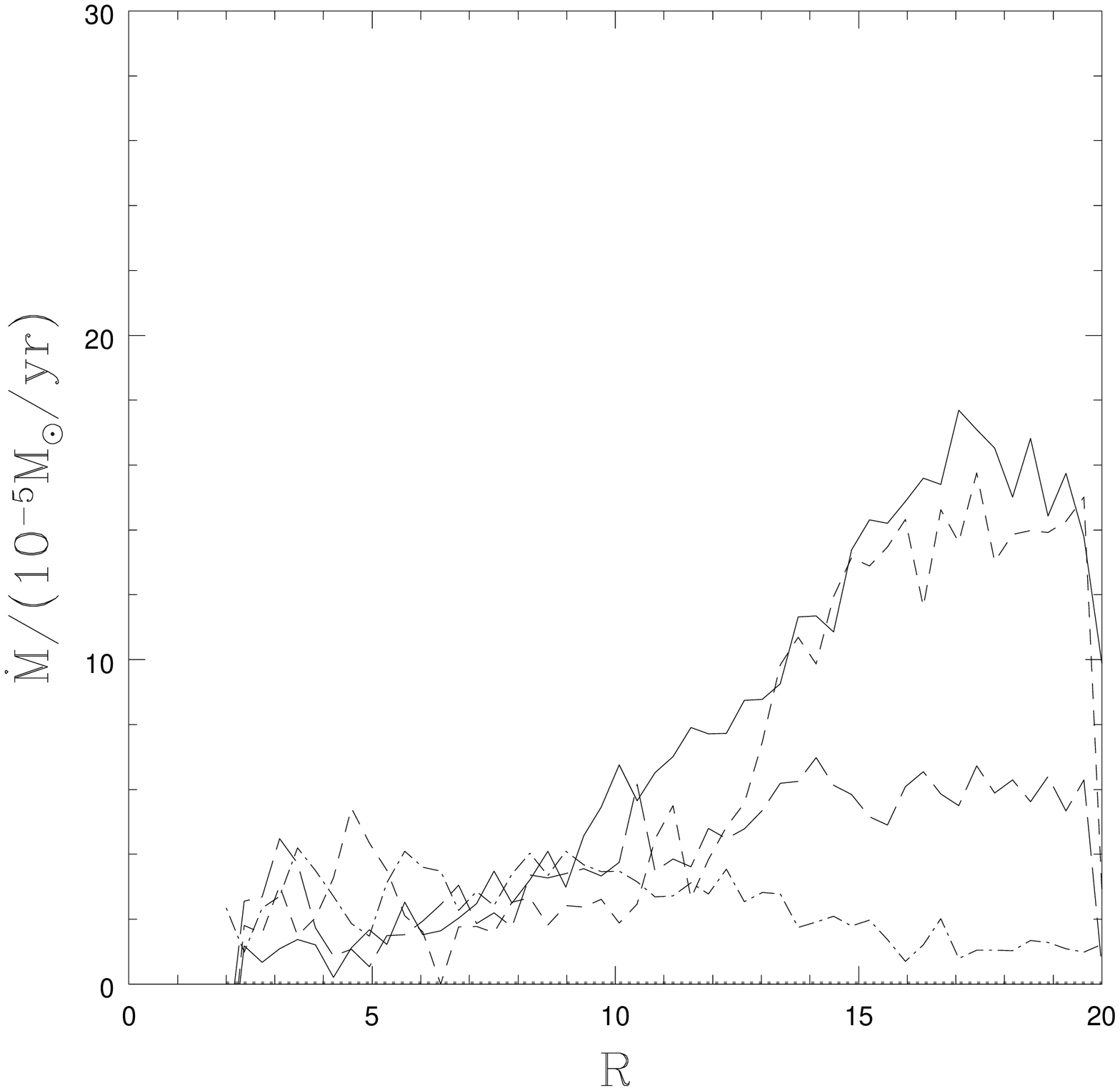,width=.45\textwidth}}
\caption{{\bf Left:} Evolution of the gravitational torque strength (as
measured by the dimensionless parameter $\alpha$, during the
transient. The four lines refer to $t=2.28t_0$ (solid line),
$t=2.52t_0$ (dashed line), $t=2.69t_0$ (long-dashed line) and
$t=2.93t_0$ (dot-dashed line). The two dotted lines show the expected
``equilibrium'' value $\alpha=0.053$, obtained from balancing viscous
heating and cooling, and the maximum torque strength ($\alpha=0.005$)
due to artificial SPH viscosity (see Paper I, Appendix). {\bf Right:}
Corresponding mass accretion rates, assuming that the central object
is $1M_{\odot}$ and that the unit length is 1 au.}
\label{fig:alphatrans}
\end{figure*}

\begin{figure}
\centerline{\epsfig{figure=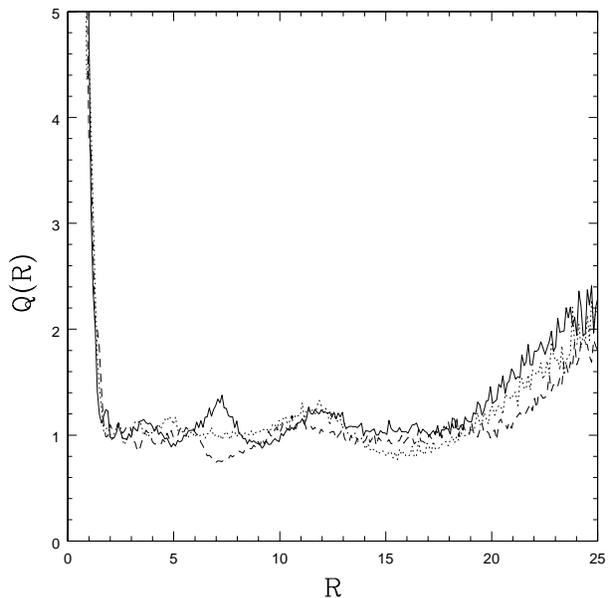,width=84mm}}
\caption{Profiles of $Q$ at $t=5.2t_0$ (solid line), $t=5.8 t_0$
(dotted line) and $t=6.6 t_0$ (dashed line).}
\label{fig:md05Q}
\end{figure}

\begin{figure}
\centerline{\psfig{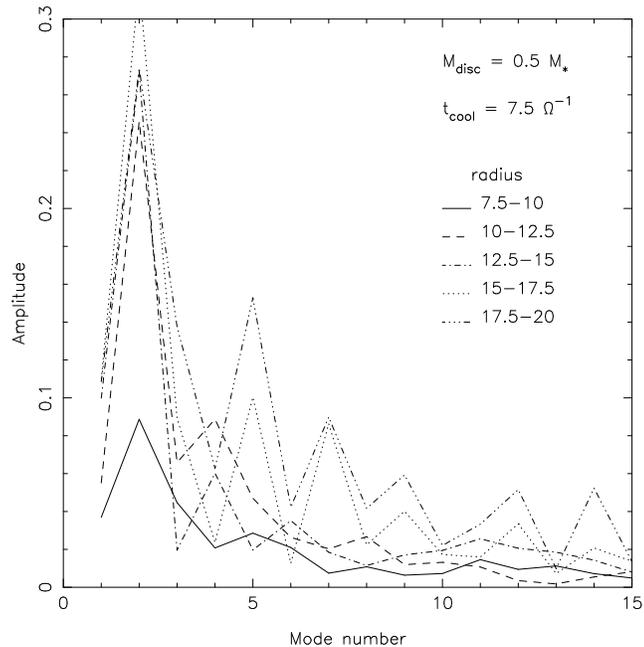}}
\caption{Amplitude of the first Fourier components of the density
structure in the $M_{\mathrm{disc}}=0.5M_*$ case for different radial
ranges in the disc.}
\label{fig:md05mode}
\end{figure}

This second episode of gravitational instability is however different
in nature to the previous one. It is not transient, but lasts for
several outer dynamical timescales (from $t\approx 4t_0$ until at least
the end of the simulation, at $t=7t_0$), showing no significant
evolution throughout. This can be seen in Fig. \ref{fig:md05Q}, which
shows the profile of $Q$ at $t=5.2t_0$ (solid line), $t=5.8 t_0$
(dotted line) and $t=6.6 t_0$ (dashed line) (cf., for comparison the
evolution shown in Fig. \ref{fig:Qtrans}).  Fig. \ref{fig:md05mode}
shows the amplitude of the first 15 Fourier components of the density
structure at the end of the simulations, computed as in Paper I. It can
be seen that the disc is dominated by low-$m$ modes, with $m\lesssim
5$, even if it does not show a clear two-armed structure as during the
first transient phase.

\begin{figure}
\centerline{\epsfig{figure=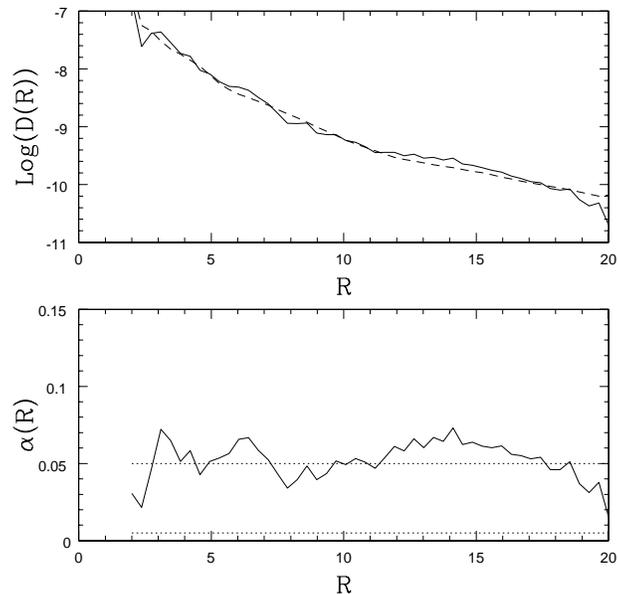,width=84mm}}
\caption{Bottom: (solid line) Effective $\alpha$ produced by
gravitational instabilities in the $M_{\mathrm{disc}}=0.5M_{\star}$
case; (dotted line) value of $\alpha$ expected from balancing viscous
heating and external cooling. The lower dotted line, shows the maximum
contribution to the stress given by artificial viscosity in SPH, as in
Fig. \ref{fig:alphatrans}. Top: (solid line) Viscous heating
power $D(R)$ compared to (dashed line) externally imposed cooling.}
\label{fig:md05alpha}
\end{figure}

\begin{figure}
\centerline{\epsfig{figure=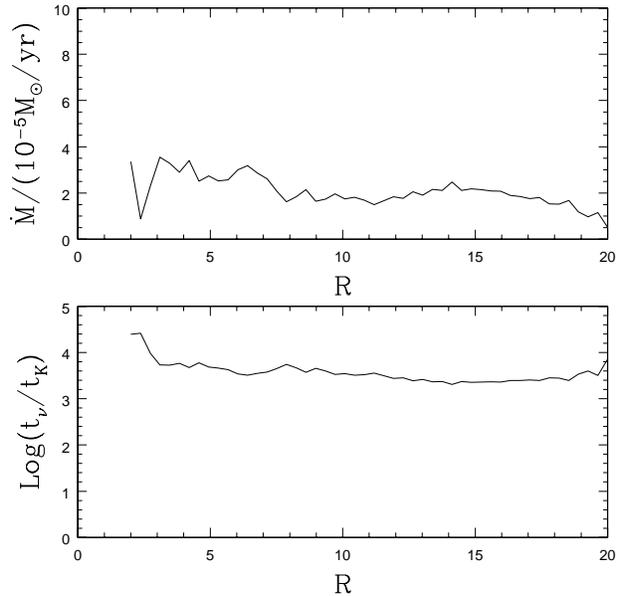,width=84mm}}
\caption{Bottom: ratio of the accretion timescale $t_{\nu}=\nu/R^2$
over the dynamical timescale $t_K=\Omega^{-1}$, for the
$M_{\mathrm{disc}}=0.5M_*$ case. Top: time averaged mass accretion rate
at the end of the simulation.}
\label{fig:md05mdot}
\end{figure}

In order to discuss the transport properties associated with this
quasi-stationary structure, we have computed the stress tensor
averaged over 640 time units, i.e $\approx 0.8t_0$. The results are
shown in Figs. \ref{fig:md05alpha} and \ref{fig:md05mdot}.  The bottom
panel of Fig. \ref{fig:md05alpha} shows the average stress tensor, as
measured by the $\alpha$ parameter (solid line), and the expected
value of $\alpha$ obtained by imposing that the viscous heating rate
exactly balances the cooling rate (dotted line). There is substantial
agreement between the two values. The lower dotted line shows, as in
the left panel of Fig. \ref{fig:alphatrans}, the maximum contribution
to the stress coming from artificial SPH viscosity. The upper panel of
Fig. \ref{fig:md05alpha} shows a comparison between the cooling rate
(dashed line) and $D(R)$, the viscous heating rate (solid line),
defined in Eq. (\ref{eq:dissipation}). Again, there is a substantial
agreement between the two suggesting that, even for this relatively
massive disc, a state of thermal equilibrium has been reached and that
the energy dissipation process is well described by a simple viscous
model.

In Fig. \ref{fig:md05mdot} we plot the radial profiles of $\dot{M}$
(upper panel) and of $t_{\nu}/t_{\mathrm{K}}$ (lower panel), where
$t_{\nu}=R^2/\nu$ is the viscous timescale and $t_{\mathrm{K}}
=\Omega_{\mathrm{K}}^{-1}$ is a measure of the dynamical timescale. We
see that the mass accretion rate acquires an almost constant value,
consistent with the quasi-stationary character of the spiral
structure, with $\dot{M}\approx 2~10^{-5} \msunyr$. Note that, even if
the value of $\alpha$ is not particularly high, the mass accretion
rate is conspicuous since the disc is relatively hot. However, despite
the high $\dot{M}$, the inspection of the bottom panel of
Fig. \ref{fig:md05mdot} shows that accretion occurs on a long
timescale, more than three orders of magnitude larger than the
dynamical timescale.

\subsection{The $M_{\mathrm{disc}}=1M_{\star}$ case}

In this case, the initial evolution of the disc is similar to that
observed in the other simulations. The disc initially simply cools down
until $Q$ becomes of the order of unity in the inner disc (where the
cooling time is fastest) and gravitational instabilities set
in. However, the long-term evolution of this massive disc differs from
the long-term evolution of the lower mass discs. A relatively
small-amplitude spiral structure, confined within $R\approx 15$, is
always present once the disc has cooled down sufficiently. However, a
series of recurrent, prominent large-scale spiral disturbances occur
repeatedly throughout the course of the simulation. The amplitude of
the $m=2$ Fourier component, in the outer disc, is plotted as a
function of time in Fig. \ref{fig:modetime}. Major episodes of spiral
activity can be easily recognized at $t\approx1.5t_0$, $t\approx 4t_0$
and $t\approx 6t_0$. A similar cyclic behaviour, with a modulation of
the amplitude of the dominant $m=2$ mode, has been found in previous
simulations of self-gravitating discs
\citep{sellwood84,laughlin94,laughlin98}.  \citet{laughlin98} have
attributed this behaviour to the effect of non-linear coupling between
different modes.

\subsubsection{The large-scale spiral}

\begin{figure}
\centerline{\psfig{figure=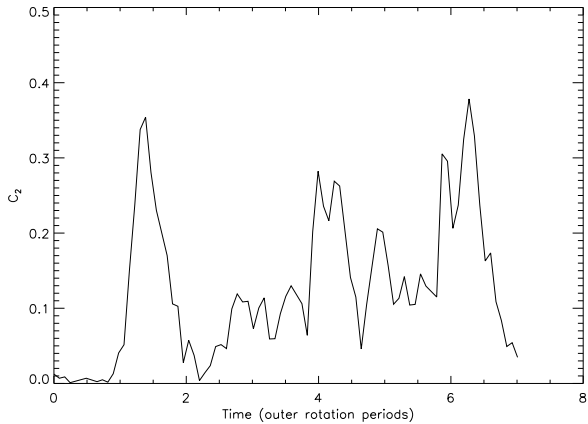,width=84mm}}
\caption{The $M_{\mathrm{disc}}=M_*$ simulation. Amplitude of the $m=2$
Fourier component of the disc surface density as a function of
time. Major episodes of spiral activity can be recognized at
$t\approx1.5t_0$, $t\approx 4t_0$ and $t\approx 6t_0$.}
\label{fig:modetime}
\end{figure}

\begin{figure*}
\centerline{\epsfig{figure=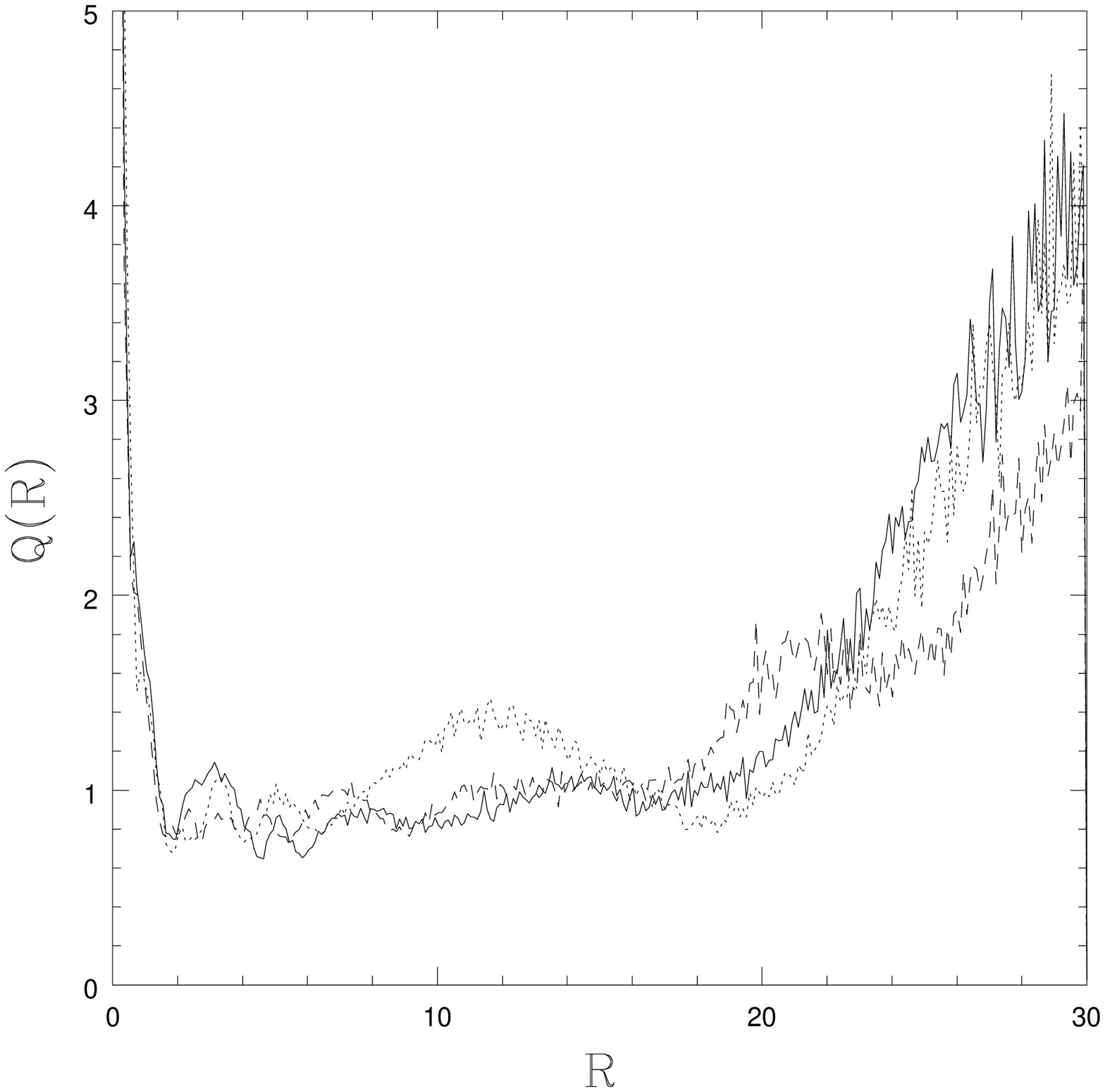,width=84mm}
            \epsfig{figure=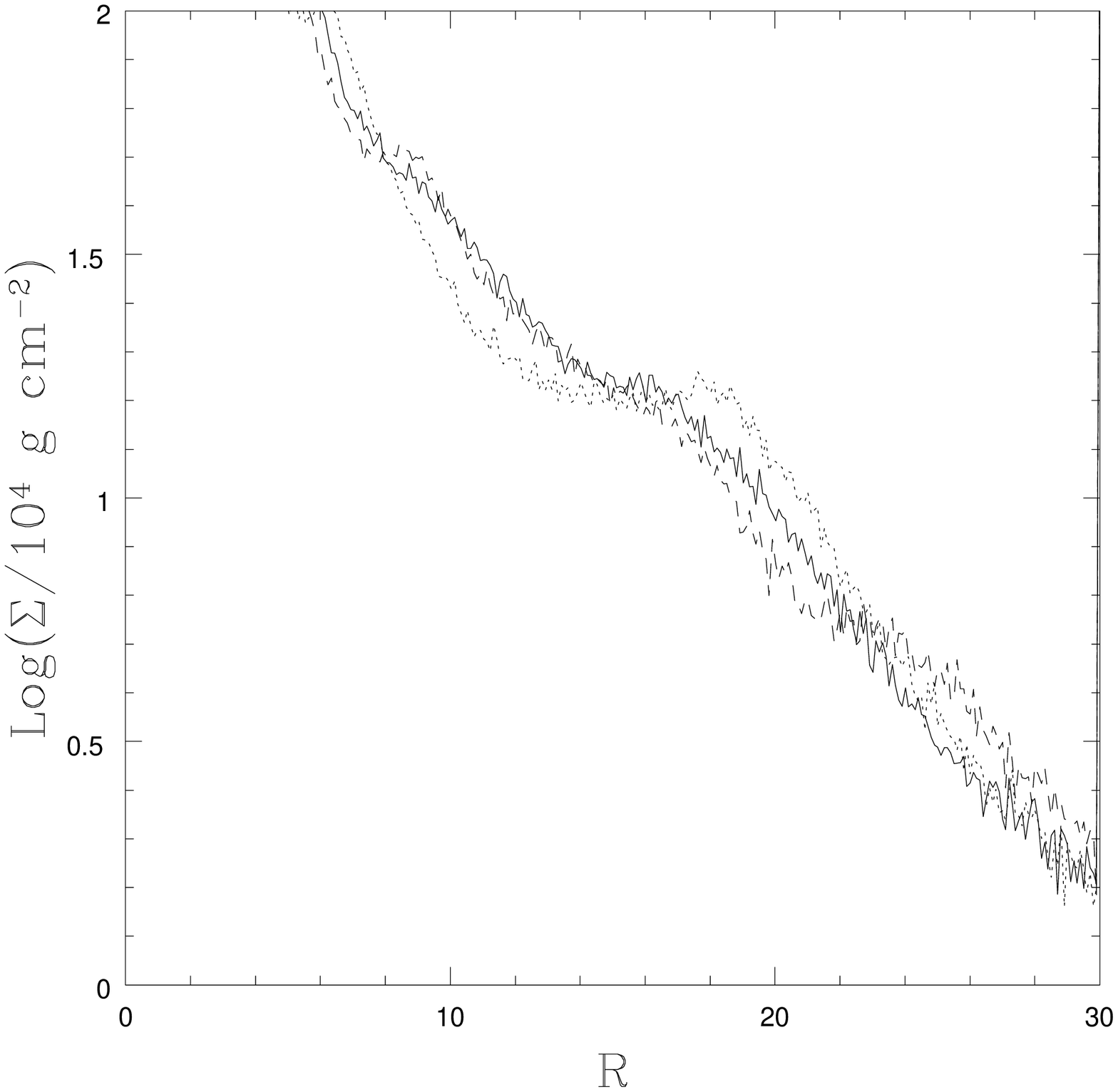,width=84mm}}
\caption{Evolution of $Q$ (left) and of the azimuthally averaged
surface density $\Sigma$ (right) during the development of a spiral
disturbance in the $M_{\mathrm{disc}} = M_{\star}$ case. The curves
refer to $t\approx 5.4 t_0$ (before the onset of the instability, solid
line), $t\approx 5.9 t_0$ (during the disturbance, dotted line), and at
$t\approx 6.3 t_0$ (after the spiral episode has vanished, dashed
line).}
\label{fig:Md1Q}
\end{figure*}

Since all the large-scale spiral disturbances share a similar physical
appearance, we will describe, as an example, one of such episodes,
which developed at $t\approx 5.8t_0$ and has vanished by $t\approx
6.5t_0$. The structure of the disc at $t\approx 5.9t_0$ is shown in
the left panel of Fig. \ref{fig:md1image} (for comparison, the right
panel shows the structure at $t\approx 3t_0$, during a period of low
spiral activity, see Fig. \ref{fig:modetime}). The disc is clearly
dominated by a large-scale $m=2$ mode, with a pattern frequency
$\Omega_{\mathrm{p}}\approx 3\Omega_{\mathrm{out}}$, resulting in
corotation at $R\approx 14$. Fig. \ref{fig:Md1Q} shows the evolution
of the profile of $Q$ (left) and of the azimuthally averaged surface
density $\Sigma$ (right) during the development of the spiral
structure. The curves refer to $t\approx 5.4 t_0$ (before the onset of
the instability, solid line), $t\approx 5.9 t_0$ (during the
disturbance, dotted line), and at $t\approx 6.5 t_0$ (after the spiral
episode has vanished, dashed line). Before the spiral develops, the
disc is characterized by a flat, self-regulated, $Q$ profile out to
$R\approx 17$. Therefore, it appears that the corotation of the spiral
mode occurs close to the outer edge of the self-regulated portion of
the disc. Note that the inner Lindblad resonance (ILR) for this mode
occurs at $R\approx 5$, and does not influence the propagation of the
density wave to small radii, in our hydrodynamical simulations. On the
other hand, it is well known that, for a stellar (collisionless) disc,
waves are efficiently absorbed at the ILR.

The effect of the development of the spiral is to remove angular
momentum from inside corotation and to release it to the matter located
outside corotation. As a result, outside corotation the disc expands
and its surface density decreases. This is evidenced from the
appearance of a local minimum in the surface density at corotation
(dotted line in the right panel of Fig. \ref{fig:Md1Q}), which
corresponds to a local maximum in the $Q$ profile. After the spiral has
vanished, the $Q$ profile is very similar to before the onset of the
transient structure, with the self-regulated part of the disc extending
out to a slightly smaller radius.

\subsubsection{Secular evolution}

\begin{figure}
\centerline{\epsfig{figure=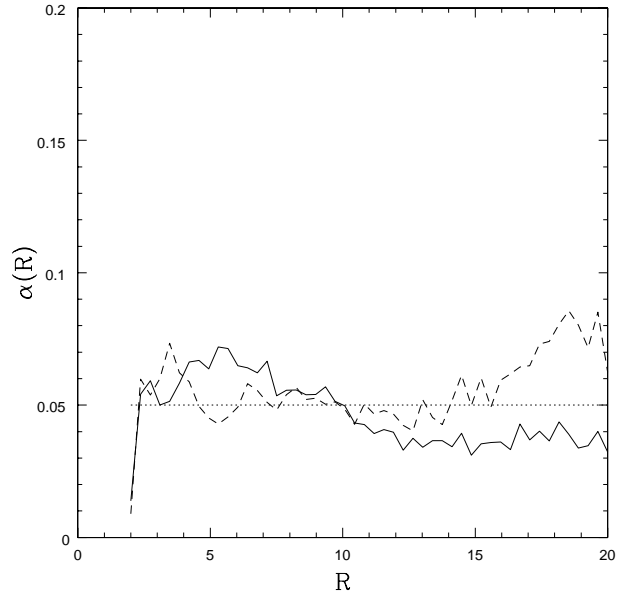,width=84mm}}
\caption{Time average of the effective $\alpha$ produced by
gravitational instabilities between $t=2.7t_0$ and $t=3.5t_0$ (where no
large scale structure is observed; solid line) and between $t=5.9t_0$
and $t=6.6t_0$ (during the development of a large scale $m=2$
disturbance, dashed line).}
\label{fig:alphahigh}
\end{figure}

As mentioned above, the appearance of a grand-design $m=2$ mode gives
rise to a substantial angular momentum transport over a small period of
time (namely, the duration of the $m=2$ disturbance). We have computed
the time average of the the effective $\alpha$ produced by the
gravitational instabilities at two different times during the evolution
of the disc: ({\it i}) between $t=2.7t_0$ and $3.5t_0$, during which
period no dominant large scale structure was observed, and ({\it ii})
between $t=5.8t_0$ and $6.6t_0$, i.e. during the development of a large
scale spiral structure (see also Fig. \ref{fig:modetime}). The results
are shown in Fig. \ref{fig:alphahigh} where $\alpha(R)$ is shown during
low spiral activity (solid line) and high spiral activity (dashed
line). During high spiral activity the gravitational torques are
larger, but not by a particularly significant amount. During both
periods the value of $\alpha$ lies very close to the expectation based
on a balance between viscous heating and cooling. If the reference
numerical values for the relevant scales are used ($1M_{\odot}$ for the
unit mass and 1 au for the unit length), the accretion rate would be of
the order of $\dot{M}\approx 10^{-4}\msunyr$. The accretion timescale
is still very large compared to the dynamical time
($t_{\nu}/t_{\mathrm{K}}\approx 10^3$), and has a very weak dependence
on radius, as in the case where $M_{\mathrm{disc}}=0.5M_{\star}$.

\begin{figure*}
\centerline{\epsfig{figure=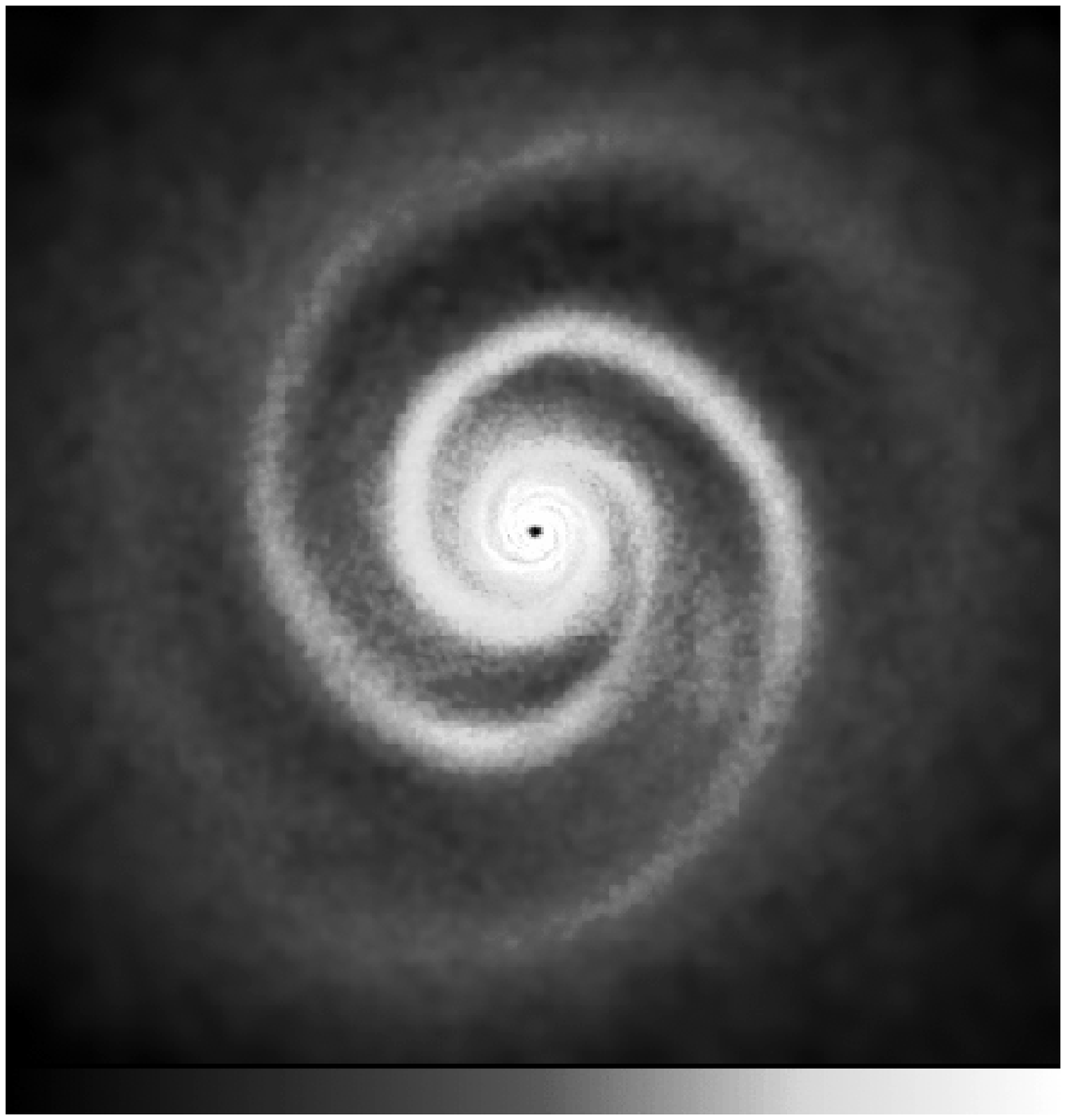,width=84mm}
            \epsfig{figure=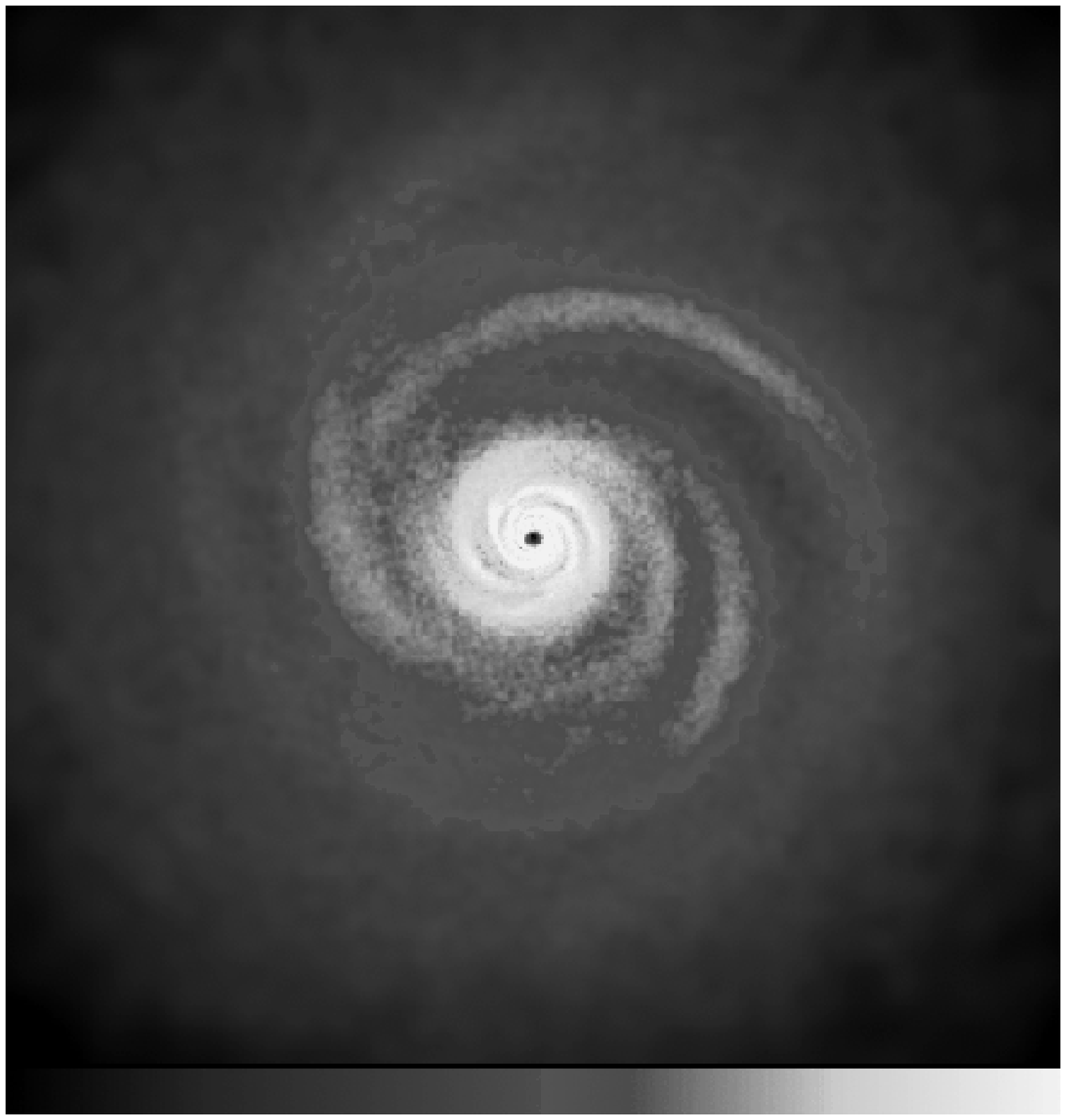,width=84mm}}
\caption{Surface density structure during the development of a spiral
instability in the $M_{\mathrm{disc}} = M_{\star}$ case, at $t\approx
5.9t_0$ (left) and at $t\approx 3t_0$ (right). Both the density and
the linear scales are the same as in Figs. \ref{fig:transient} and 
\ref{fig:md05image}.}
\label{fig:md1image}
\end{figure*}

It is interesting to compare the torque computed from
Eq. (\ref{eq:torque_alpha}), as shown in Fig. \ref{fig:alphahigh}, with
the expectations based on the linear modal theory of spiral structure
in galaxy discs. The angular momentum flux associated with a given
spiral mode can be computed based on the relevant wave action and group
velocity (see \citealt{bertin83}). We can thus express the angular
momentum flux in terms of an effective $\alpha_\mathrm{m}$, associated
with a mode with azimuthal wave number $m$. For a tightly wound mode in
a light disc we have:
\begin{equation}
\alpha_m=\left|\frac{\de\ln\Omega}{\de\ln R}\right|^{-1}
 m\epsilon_0\frac{2}{Q^2\hat{k}}
\left(1-\frac{Q^2\hat{k}}{2}\right)|\Delta|^2,
\end{equation}
where $\Delta=\delta\Sigma_\mathrm{m}/\Sigma$ is the amplitude of the
mode, $\epsilon_0=\pi G\Sigma/R\kappa^2$ and $\hat{k}=2\epsilon_0Rk$ is
the dimensionless radial wavenumber of the mode. If we measure the
amplitude $\Delta$ for the $m=2$ spiral mode shown in the left panel of
Fig. \ref{fig:md1image}, and consider the short trailing branch of the
dispersion relation, outside corotation we obtain $\alpha_{m=2}\approx
0.03$, which shows that a significant fraction of the angular momentum
transport is indeed associated with the global $m=2$ mode. Of course
some effects beyond the simple linear theory for tightly wound modes in
light discs, used in the above estimate, are expected (and should be
further analyzed) in the present case, in which the structure has
reached non-linear amplitudes and develops in a relatively heavy disc.

An important question to be addressed is whether energy dissipation
during the high spiral activity is intrinsically different to standard
viscous dissipation, or, in other words, whether wave transport of
energy is significant in our simulations. As discussed in section
\ref{sec:basic} the test that we have generally used in order to
assess this issue is a comparison between the value of $\alpha$ as
computed from the gravitational torque strength and the expected value
obtained from a balance between viscous heating and cooling. In all
simulations up to now we have interpreted the agreement between the
two values as a demonstration that wave transport does not play a
major role. In the present case, with $M_{\mathrm{disc}}=M_{\star}$,
gravitational instabilities are clearly global and different in nature
with respect to the lower mass cases: here the pattern of the
instability rapidly changes with time (in particular, the strength if
the dominant $m=2$ mode, see Fig. \ref{fig:modetime}) and the value of
$Q$, although always relatively close to unity, never really settles
down in a self-regulated state. From Fig. \ref{fig:alphahigh} we see
that even during the high spiral activity, the value of $\alpha$ is
relatively close to the expected value (dashed line in
Fig. \ref{fig:alphahigh}), being only slightly larger in the outer
disc. It would be, however, wrong to interpret this small disagreement
as an indication that wave energy transport is dominant, since the
agreement is only expected if the disc is in thermal equilibrium. The
change in internal energy of the disc as a result of the heating due
to the development of the spiral can easily account for the
discrepancy shown in Fig. \ref{fig:alphahigh}.

These results do, however, suggest that massive discs never settle into
a state of thermal equilibrium. If we interpret this in the same way as
\citep{laughlin98}, non-linear coupling between the different modes results 
in episodes when $m=2$ modes heat the outer disc and make it more stable,
followed be episodes when cooling dominates, and the disc is driven
back towards a period of instability.

\section{Conclusions}
\label{sec:conclusions}

In this paper we have extended our previous analysis of the transport
properties of self-gravitating discs, presented in Paper I. In
particular, here we describe the results of two new simulations, in
which the disc mass is of the same order of the central object mass,
namely $M_{\mathrm{disc}}=0.5M_{\star}$ and $M_{\mathrm{disc}}=
1M_{\star}$. Given the specific region of the parameter space that we
explore, our simulations have two previous analogues: the $N$-body
simulations by \citet{sellwood84}, who mimicked the cooling of the
stellar component of a galaxy disc through continuous accretion of
stars, and a series of simulations by Laughlin et al.
\citep{laughlin94,laughlin97,laughlin98}, who simulated massive gaseous
disc, adopting simple equations of state (mostly isothermal) without
any specific cooling. The main difference between the latter
simulations and those presented here lies in the way we handle the disc
thermodynamics. In fact, we explicitly solve the energy equation of the
disc, and introduce an external cooling term, simply parameterized in
terms of a cooling time $t_{\mathrm{cool}}=\beta\Omega^{-1}$. By
suitably choosing the parameter $\beta$, we are able to prevent the
disc from fragmenting, and to produce a persistent spiral structure.

The specific aim of this work is to study the transport of angular
momentum induced by gravitational instabilities and the associated
energy dissipation process, to test in particular whether wave
transport of energy plays an important role in the disc dynamics
\citep{balbus99}. In this respect, the results of this paper can be
summarized as follows:

\begin{enumerate}
\item[(i)] The development of a gravitational spiral structure is
indeed able to transport efficiently angular momentum in the disc,
hence favouring accretion. The time-averaged strength of the the
stress induced by the disc self-gravity, as measured by the parameter
$\alpha$ \citep{shakura73}, is $\alpha\approx 0.05$, so that the
associated energy dissipation (computed from
Eq. (\ref{eq:dissipation})) almost balances the externally imposed
cooling. This shows that global wave transport does not play a major
role in our simulations, in line with the predictions of
\citet{balbus99}, who argued that self-regulated discs, with $Q\approx
1$ (as indeed are ours) should be free from wave transport effects.
\item[(ii)] Contrary to the results of \citet{laughlin94}, the
accretion process induced by the disc self-gravity occurs on the long
'viscous' time-scale, rather than on the much shorter dynamical
time-scale, so that a massive self-gravitating disc can survive for a
relatively long period (see Fig. \ref{fig:md05mdot}). This important
difference can be understood by simply considering the parameter $Q$,
as defined in Eq. (\ref{eq:q}). If $Q$ is smaller than unity,
gravitational instabilities develop and the disc responds so as to
increase the value of $Q$ to reach a stable configuration. If the disc
is isothermal (as in \citealt{laughlin94}), the only way this can be
achieved is by significantly reducing $\Sigma$, i.e. by inducing a fast
accretion process. On the other hand, if the disc is allowed to
dissipate energy and to heat up, increasing the value of
$c_{\mathrm{s}}$ (as in our simulations), the disc is stabilized more
easily and the required amplitude of the gravitational disturbance is
smaller, hence producing a slower accretion.
\item[(iii)] When the disc mass is small with respect to the central
object, as in the case discussed in Paper I, the disc rapidly settles
down is a self-regulated state where the amplitude of the spiral
structure and the radial profile of $Q$ change very little with
time. In these cases, the spiral structure is tightly wound and
dominated by high-$m$ modes. In the high disc mass case presented here,
the evolution is more complex. In the case where
$M_{\mathrm{disc}}=0.5M_{\star}$ we observe the development of an
initial transient $m=2$ spiral structure, which induces a strong
redistribution of matter and a steepening of the surface density
profile, before the disc is eventually able to settle down as in the
low disc mass case. Similar initial transients are also observed by
\citet{mejia04}. We attribute it to an effect of the initial condition
adopted. In particular, the chosen initial surface density profile
$\Sigma\propto R^{-1}$ is actually shallower than analytical estimates
of the surface density of self-regulated massive discs \citep{BL99}. We
have indeed shown that the initial transient steepens the surface
density profile in such a way that it matches the analytical
estimates. The $M_{\mathrm{disc}}=1M_{\star}$ is even more complex: the
disc never really settles down and it is self-regulated only in a
time-averaged sense. The amplitude of the spiral disturbances changes
strongly with time, and recurrent, short-lived, low-$m$ spiral patterns
develop in the disc (see Fig. \ref{fig:modetime}). A similar behaviour
has been also observed in the cooling simulations by \citet{sellwood84}
and \citet{laughlin98}.
\end{enumerate}

Another interesting results of our simulations is that, even at the
high mass ratios considered here, they appear to be still consistent
with the fragmentation criterion $t_{\mathrm{cool}}<3\Omega^{-1}$. The
above criterion had been previously tested either in local
\citep{gammie01}, or in global, low disc mass simulations
\citep{rice03c}. Even if the global simulations by \citet{rice03c} show
that the actual threshold value for the cooling time is slightly
increased with respect to the local estimates by \citet{gammie01}, we
have shown here that up to $M_{\mathrm{disc}}=1M_{\star}$, the
condition $t_{\mathrm{cool}}> 7.5 \Omega^{-1}$ effectively prevents the
fragmentation of the disc. In fact, note that, even if rather large
density perturbations are experienced by the disc, especially during
the initial transients, none of these density enhancements is
long-lived. This is due primarily to our choice for the cooling time
scale and not to a limited resolution in our code. Indeed, simulations
performed with the same number of particles as ours, but with a shorter
cooling time, did show effective fragmentation \citep{rice03c}. 

The present work completes the analysis described in Paper I, but can
still be subject to many further refinements. In particular, all the
results presented here and in Paper I are based on a specific form of
the cooling term in the energy equation. On the other hand, as already
mentioned, the outcome of gravitational instabilities in discs strongly
depends on the assumptions about the disc thermal behaviour.
Simulations that employ a cooling time constant with radius seem to
show significant global energy transport \citep{mejia04}, in contrast
to our results. Of course, both a constant cooling timescale and our
choice are highly idealized assumptions, and further studies are needed
to assess which of the two is more reasonable. In principle, one could
follow \citet{gammie03} and find a relation between cooling time and
density, based on some opacity prescription. This would not, in
general, give $t_{\mathrm{cool}}\Omega=$const. However, a cooling time
proportional to the dynamical timescale has the attractive feature of
leading to a constant $\alpha$, and since self-gravitating discs
naturally evolve towards a self-regulated state, this might not be an
unreasonable assumption (see also \citealt{bertin97}). 

\begin{figure}
\centerline{\epsfig{figure=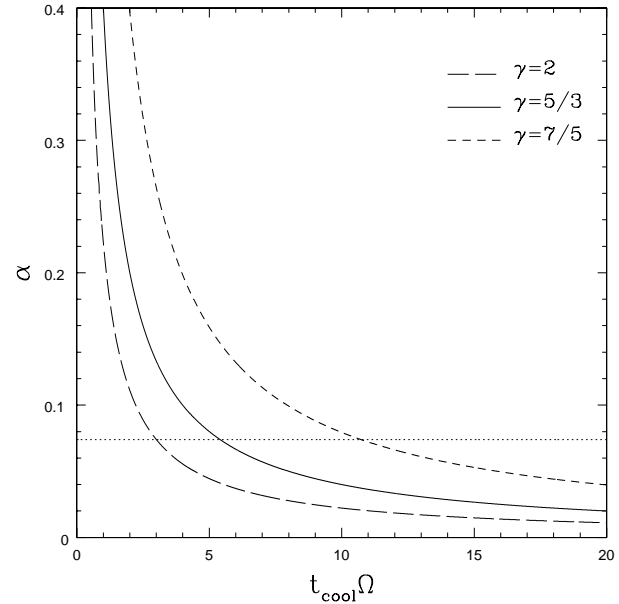,width=84mm}}
\caption{Relationship between $\alpha$ and $t_{\mathrm{cool}}$,
described in Eq. (\ref{eq:gammie}), for different values of $\gamma$:
$\gamma=2$ (long-dashed line), $\gamma=5/3$ (solid line), $\gamma=7/5$
(short-dashed line). The dotted line indicates the value of $\alpha$
corresponding to $t_{\mathrm{cool}}=3\Omega^{-1}$ in the case where
$\gamma=2$ (the value adopted by \citealt{gammie01}).}
\label{fig:alpha_cool}
\end{figure}

Another important issue to be explored is how the disc response changes
when different values for the ratio of the specific heats $\gamma$ is
used and whether this can affect the fragmentation criterion.
Preliminary tests (Rice \& Lodato, in preparation) seem to indicate
that, when $\gamma$ is decreased, the boundary cooling time for
fragmentation increases. Actually, it might be more reasonable to
express the fragmentation criterion in terms of a maximum gravitational
stress sustainable by a non-fragmenting disc. Fig. \ref{fig:alpha_cool}
shows the relationship between $\alpha$ and $t_{\mathrm{cool}}$ implied
by Eq. (\ref{eq:gammie}), for different values of $\gamma$. The dotted
line indicates the value of $\alpha$ corresponding to
$t_{\mathrm{cool}}=3\Omega^{-1}$ when $\gamma=2$ (the value adopted by
\citealt{gammie01}). In Paper I and here, we have shown that, when the
disc does not fragment, the strength of the gravitational disturbances
is such that Eq. (\ref{eq:gammie}) is approximately satisfied. If
indeed there is a maximum $\alpha$ sustainable by a self-gravitating
disc, then the disc would fragment whenever Eq. (\ref{eq:gammie})
implies a larger value of $\alpha$. This would then be consistent with
an increase of the limiting cooling time when $\gamma$ is decreased
(see Fig. \ref{fig:alpha_cool}). Further investigations are clearly
needed to asses whether this is the case or not.

\section*{acknowledgements}
The simulations presented in this work were performed using the UK
Astrophysical Fluid Facility (UKAFF). GL acknowledges support from the
EU Research Training Network {\it Young Stellar Clusters}. WKMR
ackowledges support from a UKAFF Fellowship. We thank Phil Armitage,
Giuseppe Bertin, Cathie Clarke and Jim Pringle for interesting
insightful discussions that significantly improved this paper.

\bibliographystyle{mn2e} 

\bibliography{lodato}

\end{document}